# Fractal Photonic Crystal Waveguides

Juan A. Monsoriu[1*], Carlos J. Zapata-Rodríguez[2],

Enrique Silvestre[2], and Walter D. Furlan[2]

[1]*Departamento de Física Aplicada, Universidad Politécnica de Valencia, E-46022 Valencia, Spain*

[2]*Departamento de Óptica, Universidad de Valencia, E-46100 Burjasstot (Valencia), Spain*

**Abstract**

We propose a new class of one-dimensional (1D) photonic waveguides: the fractal photonic crystal waveguides (FPCWs). These structures are photonic crystal waveguides (PCWs) etched with fratal distribution of grooves such as Cantor bars. The transmission properties of the FPCWs are investigated and compared with those of the conventional 1D PCWs. It is shown that the FPCW transmission spectrum has self-similarity properties associated with the fractal distribution of grooves. Furthermore, FPCWs exhibit sharp localized transmissions peaks that are approximately equidistant inside the photonic band gap.

---

[*] *E-mail address*: jmonsori@fis.upv.es

# 1. Introduction

In recent years the study of fractals has attracted the attention of researchers, encouraged by the fact that many physical phenomena, natural structures and statistical processes can be analyzed and described by using a fractal approach [1]. From a mathematical point of view the concept of fractal is associated with a geometrical object which 1) is self-similar (i.e., the object is exactly or approximately similar to a part of itself) and, 2) has a fractional (or noninteger) dimension. Self-similar structures are obtained by performing a basic operation, called *generator*, on a given geometrical object called *initiator*, and repeating this process on multiple levels, in each one of them an object composed of sub-units of itself is created that resembles the structure of the whole object. Mathematically, this property should hold on all (infinite) scales. However, in the real world, there are necessarily lower and upper bounds over which such self-similar behaviour applies.

In optics, fractal structures, ranging from simple one-dimensional (1D) objects [2] to complex 2D systems [3-5] have been extensively studied. Specifically, 1D dielectric multilayers characterized by a refractive-index distribution that follows Cantor [6,7] sequences have been analyzed. The fractal profile of these structures leads to a transmission spectrum with self-scaling properties [8], i.e., the variation of the spectrum with the frequency at each higher stage is a modulated version of that associated with the previous step with an appropriate scaling of the frequency range [9,10]. Moreover, as a consequence of this fractal property, these structures exhibit a certain number of transmission peaks inside the frequency band gap. The same phenomenon arises in corrugated waveguides, which film thickness follows a Cantor distribution. These systems have been analyzed using a transmission lines method based on an equivalent 1D multilayer of effective indices [11].

In this work we present a novel family of fractal photonic crystal waveguides (FPCWs) with interesting fractal properties. These systems can be generated by etching a set of grooves in a slab waveguide according with a self-similar distribution. Although there is a great deal of interest in the properties of conventional PCWs [12] to control the emission and propagation of light [13], to our knowledge, self-similar distributions of grooves in PCWs have not been studied in the literature. To analyze these systems we will employ a rigorous and efficient technique for modeling dielectric systems with an arbitrary refractive index distribution [14], since the transmission properties of 1D periodic PCW or FPCWs may differ substantially from those of ideal 1D systems.

To analyze a FPCW we will consider this system as a conventional PCW with a periodic lattice of grooves where the separation between specific grooves is increased. The resultant structure can be interpreted as a quasi-periodic PCW, i.e., a PCW with defects. The insertion of defects in conventional PCWs has been used to increase the functionally of this systems [15,16].

This paper is organized as follows. In Section 2 we revise a conventional 1D PCW and we obtain the limits of the photonic band gap and the transmission spectrum of the structure. Next, in Section 3 we tackle the problem of the synthesis of FPCWs. We perform the analysis for a particular case and we compare the transmission spectrum with the conventional PCW. Finally, in Section 4 of this paper the main results are outlined and some applications are proposed.

**2. Main aspects of the spectral transmission in conventional photonic crystal waveguides**

A conventional PCW is shown in Fig. 1. It consists of a structured thin film of refractive index $n_g$ over a substrate of index $n_s$, air being the cover. The film of thickness $d$ is etched with a finite 1D lattice of grooves of period $\Lambda$ and air-filling fraction $f$. The slab waveguide confines light in the vertical direction and the lattice of grooves allows to control the light propagating along the in-plane direction.

One of the more relevant properties of photonic crystals is the possibility to show photonic band gaps (PBGs), i.e., certain ranges of frequencies where light propagation is prohibited. In order to obtain the photonic band-gap limits of the periodic structure, we have analyzed the dispersion relation of the guided Bloch modes of the system with an infinite number of periods by using the orthonormal-basis method [17]. This technique is based on a modal algorithm that allows us to simulate dielectric systems with an arbitrary refractive-index distribution. It consists in expanding the 3D wave equation for the magnetic field in a certain domain in terms of an auxiliary system properly chosen to provide a suitable basis. In the present case, the system is periodic along the propagation direction, and therefore we must select periodic boundary conditions in that direction, and we impose confining boundary conditions in the direction perpendicular to the film plane. Then, for computational purposes, the system is placed between two parallel perfectly-conducting metallic plates located far enough from the PCW. Accordingly, the chosen auxiliary system is a 1D homogeneous air-filled waveguide, whose modes are well know.

As a practical example we considered a periodic PCW with $d = 0.8\Lambda$, $f = 0.15$, $n_g = 3.37$, and $n_s = 1.61$. These refractive-index values correspond to a thin-film of GaAs over a substrate of $Al_xO_y$. Its TE Brillouin diagram is shown in Fig. 2. The upper left and right shaded regions correspond to radiation modes. The dispersion curves allow to localize the PGB between the normalized frequency ($2\pi\lambda/\Lambda$) values 1.155 and 1.327. In this frequency range only evanescencent modes characterized by a complex wavevector $k_z$ are solutions of the wave equation and, therefore, the transmission spectrum vanishes in an infinite periodic structure. However, when the number of periods is finite, we should match these modes and those of the waveguides connected to the PCW at the input and output planes, and, consequently, part of the energy may pass through the PCW.

We have determined the PCW transmission spectrum with a finite number of periods by using the numerical method proposed in Ref. 14. This method allows to analyze 3D dielectric systems with an arbitrary refractive index distribution connected to dielectric waveguides. In our case, the 3D dielectric system corresponds to the central region of Fig. 1, where the air grooves are located, which is connected to the input and output slab waveguides. The present method provides a linear equation system relating the modal coefficients of the waveguide modes, which allow to obtain the multimode scattering matrix. These waveguide modes can be easily obtained with the biorthonormal-basis method [18,19].

In Fig. 3, we present the transmission spectrum of the conventional PCW shown in Fig. 1 for different numbers of grooves, $N$. We observe that the addition of grooves causes a rapid decrease of the transmitted power inside the PBG range. The minimum of the normalized transmitted power with only 8 air grooves is 0.049.

## 3. Self-similarity in photonic crystal waveguides

The main idea of the present work consists in introducing some defects in the conventional PCW shown in Fig. 1, such that the air grooves distribution follow a self-similar sequence. As a particular case we have selected the triadic Cantor sequence shown in Fig. 4. The first step in the construction procedure of this particular self-similar sequence consists in defining a straight-line segment of length, $L$, called the initiator (stage $S = 0$). Next, at the stage $S = 1$ the generator of the set is constructed from $N = 2$ non-overlapping copies of the initiator, each with a scale $\gamma<0.5$. At following stages, the generation processes is repeated over and over again for each segment in the previous stage. At the $S$th stage, the triadic

Cantor fractal consists of $2^S$ segments interleaved with $2^S-1$ gaps, all segments having equal lengths:

$$a^{(S)} = L\gamma^S. \tag{1}$$

The length of the $i$th gap (numbered from left to right in Fig. 4) of the $S$th stage is given by

$$b_i^{(S)} = L(1-2\gamma)\gamma^{S-k[S,i]}, \tag{2}$$

where $k[S,i]$ is an integer ($1 \leq k[S,i] \leq S$) such that $i/2^{k[S,i]-1}$ is an odd integer (hereafter $k[S,i]$ will be written as $k$ for simplicity).

According with the above self-similar sequence, at $S$th stage, the proposed FPCW consists of $2^S$ air grooves etched in a slab waveguide (see in Fig. 4). Note that this system can be considered as a quasi-periodic PCW, i.e., a conventional PCW of period

$$\Lambda^{(S)} = a^{(S)} + b_1^{(S)} = L(1-\gamma)\gamma^{S-1}, \tag{3}$$

and an air-filling fraction given by

$$f = \frac{a^{(S)}}{\Lambda^{(S)}} = \frac{\gamma}{1-\gamma}, \tag{4}$$

where a set of defects have been properly introduced by increasing the distance between specific air grooves. When $S > 1$, the total number of introduced defects coincides with the number of gaps of the previous stage, S-1, i.e., $2^{S-1}-1$. These defects are located at the $i$th gap when $k > 1$ in Eq. (2). The resulting FPCW is shown in Fig. 5 for $S = 3$.

The TE$_m$ ($m = 0,1,...$) guided modes in the slab waveguide have cut frequencies given by the expression

$$\omega_c^{(m)} = \frac{1}{d\Delta}(\Phi + m\pi) \tag{5}$$

where $\Delta = \sqrt{n_g^2 - n_s^2}$ and $\tan[\Phi] = \sqrt{n_s^2 - 1}/\Delta$. Singlemode guidance is obtained when the radiation frequency varies in the interval

$$\omega_c^{(0)} < \omega < \omega_c^{(1)}. \tag{6}$$

Since the system frequency range could not be scaled, in order to show the self-similarity properties of the FPCW spectrum for different values of $S$, we must scale the whole structure leaving in each stage the air grooves with the same width $a$. In this case, the period of FPCW is also a constant magnitude, $\Lambda$, where $a = f\Lambda$ [see Eq. (4)]. In terms of the structural parameters $\Lambda$ and $f$, the distance between air grooves in the FPCW is given by

$$b_i^{(S)} = \Lambda(1-f)\left(\frac{1+f}{f}\right)^{k-1}, \tag{7}$$

and the FPCW length is

$$L^{(S)} = \Lambda(1+f)^S f^{1-S}. \tag{8}$$

From Eq. (7), when $k = 1$ (odd values of $i$) the distance between air grooves coincides with that of a conventional PCW, $(1-f)\Lambda$. At other positions with $k > 1$ (even values of $i$), a defect has been generated in the structure increasing the distance between air grooves in a factor $[(1+f)/f]^{k-1}$.

Next we analyze a FPCW for different stages, $S$, and the structural parameters used to obtain the transmission spectrum in Fig. 3 for the conventional PCW ($d = 0.8\Lambda$, $f = 0.15$, $n_g = 3.37$, and $n_s = 1.61$). From Eqs. (5) and (6), singlemode waveguide operation is given in the range comprised between the normalized frequencies 0.170 and 1.497. We focus our attention around the PBG shown in Fig. 2 and we will compare the spectrum of the FPCW with that of the conventional PCW (Fig. 3). Fig. 6 shows the transmission spectrum of the considered FPCW for stages $S = 1, 2$ and 3. These results have been obtained using the same numerical methods implemented in the previous section. We observe that the transmission spectrum of the system exhibit a characteristic fractal profile, reproducing the self-similarity of the original FPCW. In fact, the response at a given stage, $S$, is a modulated version of that associated with the previous stage.

Obviously, when $S = 1$ the system is simply a conventional PCW with N = 2, so the resultant transmission spectrum coincides with that obtained in Fig. 3. For $S = 2$, we have a quasi-periodic structure with 4 air grooves and a central defect located at $i = 2$ ($k = 2$). As a consequence, resonant frequencies appears in the spectrum, at normalized frequency values separated approximately 0.123. For $S = 3$, the resulting quasi-periodic structure has 8 air grooves and 3 defects located at positions $i = 2$, 4 and 6 ($k = 2$, 3 and 2, respectively). New resonant frequencies arise in the PBG, characterized by narrower widths. The distance between these frequencies diminishes to about 0.016. Deviation of these values may be attributed to the fact that FPCWs are dispersive structures since the effective index of the guide mode depends on the frequency (even if the media of the slab-waveguide are not). With the considered air-filling fraction, $f = 0.15$, the scale factor of the Cantor sequence given in Eq. (4) is approximately 0.130. A simple evaluation of the fraction of resonance distances for $S = 3$ and $S = 2$ gives an identical value. This fact illustrate the self-similarity property of this kind of structures.

## 4. Conclusions

The FPCW, a new type of PCW having a self-similar structure has been introduced. The design process of these waveguides is described and the expressions for the construction parameters are derived. It is shown that a given FPCW can be understood as a conventional PCW with some defects introduced in the lattice. As a particular example, the numerical simulation of a system generated by etching a finite set of grooves in a slab waveguide distributed according with a triadic Cantor sequence has been proposed. The FPCW gives a transmission spectrum characterized by a fractal profile, reproducing the structural self-similarity of the originating system. The transmission spectra of the FPCW presents sharp localized resonances approximately equidistant inside the photonic band gap as a consequence of the fractals properties. These transmission spectra could be used to design novel polychromatic optical filters. Moreover, the field localization associated to the above resonances could be useful in designing nonlinear devices. Novel FPCW following alternative 1D and 2D self-similar sequences are subject of continuing study.

**Acknowledgments**

This research has been supported by the Plan Nacional I+D+I (grant TIC 2002-04527-C02-02), Ministerio de Ciencia y Tecnología, Spain. We also acknowledge the financial suport from the Generalitat Valenciana (grant GRUPOS03/227 and GV04B-082), Spain.

**Figure captions**

**Figure 1.** Schematic representation of a conventional PCW with a finite 1D lattice of air grooves integrated between two longitudinaly homogeneous waveguides.

**Figure 2.** TE Brillouin diagram for a conventional PCW with an infinite number of periods.

**Figure 3.** Transmission spectrum of the conventional PCW shown in Fig. 1 ($d = 0.8\Lambda$, $f = 0.15$, $n_g = 3.37$, and $n_s = 1.61$) with a finite number of periods, $N$, for TE polarization.

**Figure 4.** Triadic Cantor set for the levels $S = 0$, 1, 2, and 3. The structure for $S = 0$ is the initiator, and the one corresponding to $S = 1$ is the generator. Black regions correspond to the grooves etched in the FPCW (see Fig. 5).

**Figure 5.** Fractal photonic crystal waveguide at stage of growth $S = 3$.

**Figure 6.** Transmission spectrum of the FPCW shown in Fig. 5 ($d = 0.8\Lambda$, $f = 0.15$, $n_g = 3.37$ and $n_s = 1.61$) at different stages of growth, $S$, and TE polarization.

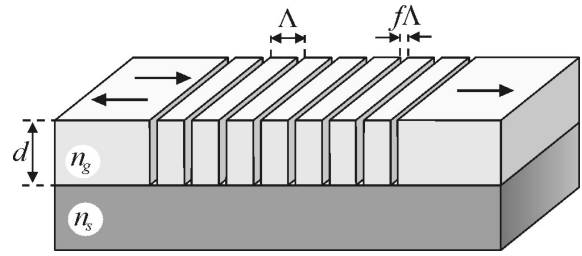

Figure 1
J.A. Monsoriu *et al.*

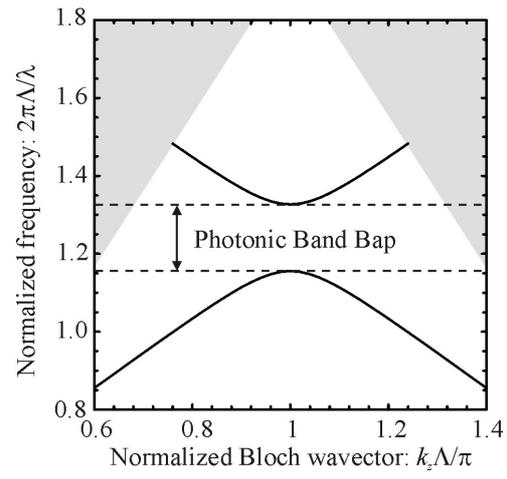

Figure 2
J.A. Monsoriu *et al.*

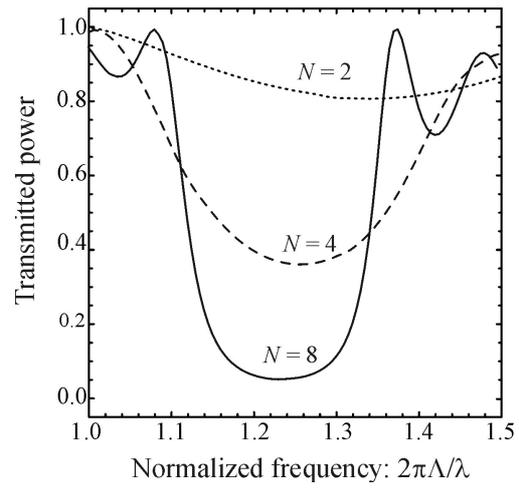

Figure 3
J.A. Monsoriu *et al.*

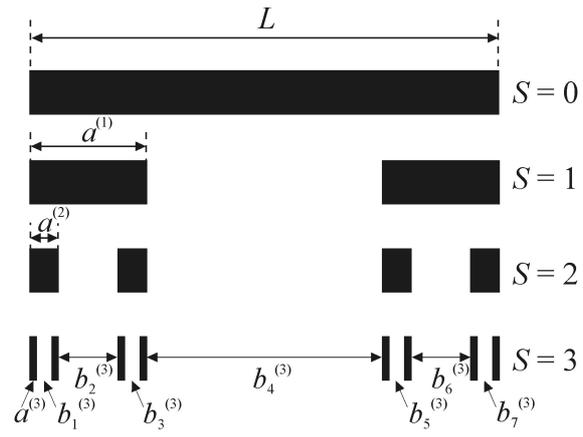

Figure 4
J.A. Monsoriu *et al.*

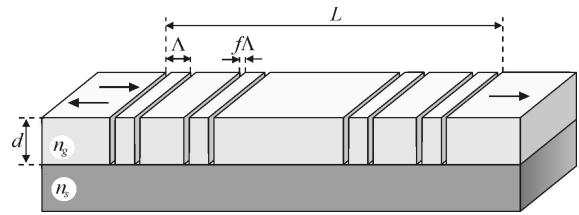

Figure 5
J.A. Monsoriu *et al.*

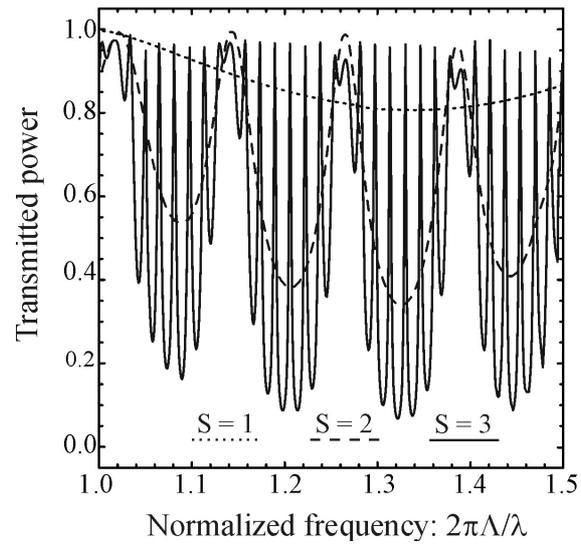

Figure 6
J.A. Monsoriu *et al.*